\documentclass[12pt]{article}
\usepackage{graphicx}
\usepackage{dcolumn}
\usepackage{bm}
\usepackage{amssymb}
\usepackage{amsmath}
\renewcommand{\(}{\hspace*{.5pt}\!\left(}
\renewcommand{\)}{\right)\hspace*{1pt}\!}

\newcommand{\bq}{\begin{equation}}
\newcommand{\eq}{\end{equation}}

\newcommand{\bqr}{\begin{eqnarray}}
\newcommand{\eqr}{\end{eqnarray}}

\newcommand{\bqrx}{\begin{eqnarray*}}
\newcommand{\eqrx}{\end{eqnarray*}}

\newcommand{\br}{\begin{array}}
\newcommand{\er}{\end{array}}

\begin{document}

\pagestyle{empty}

\setlength{\parindent}{18pt}
\setlength{\footskip}{.5in}



\vspace*{.6in}
\begin{center}
Weak limits for quantum walks on the half-line
\end{center}
\begin{center}
Chaobin\, Liu\footnote{cliu@bowiestate.edu} \quad Nelson Petulante\footnote{npetulante@bowiestate.edu\\Department of Mathematics, Bowie State University \, Bowie, MD 20715 USA\\ 
}\\
\end{center}

\begin{abstract}

For a discrete two-state quantum walk on the half-line with a general condition at the boundary, we formulate and prove a weak limit theorem describing the terminal behavior of its transition probabilities. In this context, localization is possible even for a walk predicated on the assumption of homogeneity. For the Hadamard walk on the half line, the weak limit is shown to be independent of the initial coin state and to exhibit no localization.



\end{abstract}


\section{Introduction}

The notion of quantum walks (QW) originated in 1993 with the publication of \cite{ADZ93}. Since then, inspired by the ideas in \cite{ABNVW01,AAKV01,FG98}, and propelled, in no small measure, by the hypothetical prospect of application to the development of super-efficient quantum algorithms, this field of research has flourished immensely. For a lively and informative elaboration of the history of quantum walks and their connection to quantum computing and physics, the reader is referred to \cite{K03, K06, K08, VA08, AMC2009, AVWW11, VA12, CGMV2012}. 

Among numerous other valuable results, \cite{K02, Konno05, KLS12} have formulated and proved a  weak limit theorem for the transition probabilities of discrete unitary quantum walks on the full line (dynamically restricted to integer nodes). In this work, we formulate and prove its counterpart for quantum walks on the half-line (dynamically restricted to non-negative integer nodes). Specifically, in this paper, we adopt the model of unitary discrete quantum walks on the half-line as proposed in \cite{CGMV2010}. 

For a QW of this kind, much effort has been dedicated to studying the asymptotic behavior of its site-specific return probabilities and the mass point aspect of these distributions, notably in \cite{CGMV2012, CGMV1-2012} using the CGMV method. Meanwhile, for a QW of this kind, the overall weak limit law for the probability distribution has remained relatively unexplored. Our treatment addresses this omission and emulates the approach employed in \cite{CHKS2009}.

This paper unfolds as follows. In the second section, we introduce the elements essential to understanding our adopted model of quantum walks on the half line and we define and develop the generating functions of the probability amplitudes associated therewith. In third section, we formulate and prove a weak limit formula for the class of QW's under consideration. Much of the derivations and justifications of technical details are relegated to the subsequent sections.

\section{QWs on the half line}

In this paper, we adopt the model of unitary quantum walks on the half-line as defined in \cite{CGMV2010}. In this model, the unitary shift operator $S$ is given by

\begin{eqnarray}
 S=\sum_{x=0}^{\infty}|x+1\rangle\langle x|\otimes |\uparrow\rangle\langle\uparrow|+\sum_{x=1}^{\infty}|x-1\rangle\langle x| \otimes |\downarrow \rangle \langle \downarrow|+|0\rangle\langle 0|\otimes |\uparrow\rangle\langle\downarrow|. \label{equso}
\end{eqnarray}

According to this formula, each temporal iteration of $S$ acts as follows: when located at any site $0, 1, 2, 3, ...$, an upward spin migrates one step to the right. When located at any strictly positive position $1, 2, 3, ...$, a downward spin migrates one step to the left. When located at the special site $0$, a downward spin reverses orientation and remains at site $0$, so that, in the next iteration, it migrates to site $1$.

In this model, the evolution of quantum states is governed by two unitary so-called ``coins", represented by the matrices:

\begin{equation}
 U_0= \left[\begin{array}{cc}
a_0& b_0\\
c_0& d_0
\end{array}\right],\quad
U= \left[\begin{array}{cc}
a& b\\
c& d
\end{array}\right], \label{Coin Matrices}
\end{equation}

\noindent in terms of which, the one-step transition rules can be framed by a master operator $\mathfrak{U}$, as follows. When $x\ne 0$, we have

\begin{eqnarray}
\mathfrak{U}|x\downarrow\rangle=a |x-1\downarrow\rangle+c|x+1\uparrow\rangle, \quad \mathfrak{U}|x\uparrow\rangle=b |x-1\downarrow\rangle+d|x+1\uparrow\rangle, \label{U_x}
\end{eqnarray}

\noindent while, in the special case when $x=0$, we have a general condition at the boundary  

\begin{eqnarray}
\mathfrak{U}|0\downarrow\rangle=a_0 |0\uparrow\rangle+c_0|1\uparrow\rangle, \quad \mathfrak{U}|0\uparrow\rangle=b_0 |0\uparrow\rangle+d_0|1\uparrow\rangle. \label{U_0}
\end{eqnarray}

For the sake of clarity, it is appropriate to distinguish between two kinds of quantum walks on the half line. If the quantum walk is governed everywhere by a single constant coin $U$ (in particular $U_0=U$ in Eq.(\ref{Coin Matrices})) let it be referred to as a {\it homogeneous} quantum walk. Otherwise, let the quantum walk be called {\it inhomogeneous}. For instance, the Hadamard quantum walk on the half line is the homogeneous quantum walk governed everywhere by the constant coin $U=H$, where $H$ is Hadamard matrix:

\begin{equation}
 H= \frac{\sqrt{2}}{2}\left[\begin{array}{cc}
1& 1\\
1& -1
\end{array}\right]. \label{H}
\end{equation}


As is customary in the literature, the probability amplitude of finding the walker at position $x$ at time $t$ is denoted by $\psi(x,t)=(\psi_{\downarrow}(x,t),\psi_{\uparrow}(x,t))^T$, while its generating function is denoted by $\Psi(x,z)=\sum_{t=0}^{\infty}\psi(x,t)z^t$.  

Referring to Eq. (\ref{Coin Matrices}), let the matrices $S_{0},\,\, Q_{0},\,\, P\,\, \mbox{and}\,\, Q$ be defined as follows: 

\begin{equation}
 S_0= \left[\begin{array}{cc}
0& 0\\
a_0& b_0
\end{array}\right],\,\,\,
Q_0= \left[\begin{array}{cc}
0& 0\\
c_0& d_0
\end{array}\right],\,\,\,
P= \left[\begin{array}{cc}
a& b\\
0& 0
\end{array}\right],\,\,\,
Q= \left[\begin{array}{cc}
0& 0\\
c& d
\end{array}\right]. \label{matrices}
\end{equation}

Then the one-step transition rules given by Eqs. (\ref{U_x}) and (\ref{U_0}) can be reformulated as follows in terms of the probability amplitudes: 

\begin{eqnarray}
\!\!\!\psi(0,t+1)
\!\!&=&\!\!P\psi(1,t)+S_0\psi(0,t) \nonumber\\
\!\!\!\psi(1,t+1)
\!\!&=&\!\!P\psi(2,t)+Q_0\psi(0,t) \nonumber\\
\!\!\!\psi(x,t+1)
\!\!&=&\!\!P\psi(x+1,t)+Q\psi(x-1,t) \quad \mathrm{for}\, x\ge 2 \label{recursive-amplitudes}
\end{eqnarray} 

Enlightened by the techniques employed in \cite{K02, OKAA05}, it is possible, in the present context, to derive an explicit closed formula for the generating function $\Psi(x,z)$.

{\bf Lemma 1}\, Suppose the quantum walk (as defined above) is launched from site $0$ with initial coin state $(\alpha, \beta)^T$  (i.e., $\psi(0,0)=(\alpha, \beta)^T$ and $\psi(x,0)=0$ for $x\ge 1$). Then the generating function  $\Psi(x,z)$ is given by

\begin{eqnarray}
\!\!\!\Psi_{\downarrow}(0,z)
\!\!&=&\!\!\alpha+\frac{(b+adB^r(0,z))(\alpha\Delta_0z+\alpha c_0+\beta d_0)z^2}{1-b_0z-bc_0z^2-b\Delta_0z^3-ad\Delta_0z^3B^r(0,z)-ac_0dz^2B^r(0,z)},\nonumber\\
\!\!\!\Psi_{\uparrow}(0,z)
\!\!&=&\!\!\beta+\frac{\alpha a_0z+\beta b_0z+\beta b\Delta_0z^3+\beta ad \Delta_0 z^3 B^r(0,z)}{1-b_0z-bc_0z^2-b\Delta_0z^3-ad\Delta_0z^3B^r(0,z)-ac_0dz^2B^r(0,z)},\nonumber\\
\!\!\!\Psi_{\downarrow}(x,z)
\!\!&=&\!\!\frac{(d\lambda_+/a)^{x-1}dzB^r(0,z)(\alpha\Delta_0z+\alpha c_0+\beta d_0)}{1-b_0z-bc_0z^2-b\Delta_0z^3-ad\Delta_0z^3B^r(0,z)-ac_0dz^2B^r(0,z)}\, \mathrm{for}\, x\ge 1,\nonumber\\
\!\!\!\Psi_{\uparrow}(x,z)
\!\!&=&\!\!\frac{(d\lambda_+/a)^{x-1}z(\alpha\Delta_0z+\alpha c_0+\beta d_0)}{1-b_0z-bc_0z^2-b\Delta_0z^3-ad\Delta_0z^3B^r(0,z)-ac_0dz^2B^r(0,z)} \, \mathrm{for}\, x\ge 1, \nonumber
\end{eqnarray}

\noindent where
\begin{eqnarray}
\lambda_+=\frac{1+\Delta z^2-\sqrt{\Delta^2 z^4+2\Delta(1-2|a|^2)z^2+1}}{2dz}, B^r(0,z)=\frac{\lambda_+-az}{acz},\\
\Delta_0=a_0d_0-b_0c_0,\, \mathrm{and}\,\Delta=ad-bc
\end{eqnarray}

The details involved in deriving the above formulas for $\Psi(x,z)$ are deferred to section 4.

Let $X_t$ denote the position of the walker at time $t$. According to standard quantum mechanical conventions, the probability distribution of $X_t$ is identified with $|\psi(x,t)|^2$. In the sequel, equipped with the explicit expressions for $\Psi(x,z)$, as given by Lemma 1, our main goal is to evaluate, for a QW on the half-line, the terminal behavior, in the weak limit sense, of the scaled quantity $X_t/t$ as $t\rightarrow \infty$.

\section{Weak limit for QW's on the half line}

To enable feasibility within reasonable effort, and without excessive loss of generality, let us proceed on the assumption that the determinants of the unitary coin operators in Eq. (\ref{Coin Matrices}) are equal. That is: $\Delta_0=\Delta$. 

Following the approach in \cite{CHKS2009}, the following theorem is obtained:

{\bf Theorem 1.}\, Suppose the QW on the half line (as defined above) is launched from site $0$ with the initial coin state $\alpha |\downarrow\rangle +\beta|\uparrow\rangle$, then 
\begin{eqnarray}
X_t/t\Rightarrow Y,
\end{eqnarray}
where $Y$ is the random variable with probability density 
\begin{eqnarray}
f(y)=\rho \delta(y)+h(y)\frac{|c|^2y^2 I_{(0,|a|)}(y)}{\pi (1-y^2)\sqrt{|a|-y^2}} \label{densityf(y)}
\end{eqnarray} 
and where the symbol $\Rightarrow$ denotes convergence in distribution. The symbols in the expression for $f(y)$ are defined as follows: 
\begin{eqnarray}
\!\!\!h(y)
\!\!&=&\!\!\frac{(h_6-h_8)(h_1+h_2-h_3)+h_7(h_4+h_5)}{(h_1+h_2-h_3)^2-(h_4+h_5)^2}\nonumber\\
\!\!&+&\!\!\frac{(h_6+h_8)(-h_1+h_2+h_3)+h_7(h_4-h_5)}{(-h_1+h_2+h_3)^2-(h_4-h_5)^2}   \label{h(y)1},
\end{eqnarray}
where 
\begin{eqnarray}
h_1=2|c|^2\Im{(\bar{c}\Delta^{\frac{1}{2}}-\bar{c_0}\Delta^{\frac{1}{2}})},\, h_2=(1+|c_0|^2-2\Re{(\bar{c}c_0)})|c|\sqrt{1-y^2},\nonumber\\
h_3=\Im{(\bar{c}\Delta^{\frac{1}{2}}+\bar{c}c_0^2\Delta^{-\frac{1}{2}})}(1-y^2),\, h_4=2|c|\Re{(\bar{c}\Delta^{\frac{1}{2}}-\bar{c_0}\Delta^{\frac{1}{2}})}\sqrt{|a|^2-y^2},\nonumber\\
h_5=2\Im{(\bar{c}c_0)}\sqrt{1-y^2}\sqrt{|a|^2-y^2},\, h_6=(|\alpha|^2+|\alpha c_0+\beta d_0|^2)\sqrt{1-y^2}\nonumber\\
h_7=2\Re{(\alpha \Delta^{\frac{1}{2}}\overline{\alpha c_0+\beta d_0})}\sqrt{|a|^2-y^2},\, h_8=2|c|\Im{(\alpha \Delta^{\frac{1}{2}}\overline{\alpha c_0+\beta d_0})},\label{h(y)2}
\end{eqnarray}
and where
\begin{eqnarray}
\rho=1-\int_0^{|a|}h(y)\frac{|c|^2y^2 I_{(0,|a|)}(y)}{\pi (1-y^2)\sqrt{|a|-y^2}}dy. \label{weight_delta}
\end{eqnarray}

The proof of this theorem can be found in Section 5. 

The coefficient $\rho$ of the delta function in Eq.(\ref{densityf(y)}) serves as the characteristic indicator of the effect called "localization". The model is said to exhibit localization if and only if $\rho>0$. Quantitatively $\rho$ equates to the sum of all limiting probabilities of finding the walker at each site.

\vskip 0.1 in

In \cite{K02, Konno05}, the author offers weak limit theorems for homogeneous quantum walks on the full line. More recently, these results have been extended to topological settings similar \cite{KLS12}, but not quite equivalent, to the half-line setting of Theorem 1. In our setting, when restricted to the special case of a homogeneous quantum walk on the half line, Theorem 1 implies the following corollary:

{\bf Corollary 1.}\, Let the homogeneous QW be launched from site $0$ with the initial coin state $\alpha |\downarrow\rangle +\beta|\uparrow\rangle$. Then
\begin{eqnarray}
X_t/t\Rightarrow Y,
\end{eqnarray}
where $Y$ is the random variable with probability density 
$$f(y)=\rho \delta(y)+\frac{2|c|^3[|\alpha|^2+|\alpha c+\beta d|^2-2\Im{(\bar{c}\Delta^{\frac{1}{2}})}\Im{(\alpha \Delta^{\frac{1}{2}}\overline{\alpha c+\beta d})}]y^2}{\pi |a|^2[|c|^2-(\Im{(\bar{c}\Delta^{\frac{1}{2}})})^2(1-y^2)](1-y^2)\sqrt{|a|^2-y^2}}I_{(0,|a|)}(y)$$

In all cases studied in \cite{K02, Konno05}, involving a two-state homogeneous QW on the full line, the phenomenon of localization never occurs as a feature of the weak limit. However, for a typical two-state homogeneous QW on the half-line, the limiting behavior is quite different. Frequently, but not always, localization is to be expected. However, this discovery is not without precedent. \cite{CGMV2012,CGMV1-2012} have devised special techniques known as the CGMV method for treating questions about mass point and the return probabilities of QWs on the half line. Also the authors (see page $1159$ in \cite{CGMV2012}) have offered an example of a homogeneous QW on the half line for which localization occurs. Here we offer a different example.

{\bf Example 1. }\,Suppose the QW, everywhere governed by the coin operator   
\begin{eqnarray}
U= \frac{\sqrt{2}}{2}\left[\begin{array}{cc}
1& e^{i\pi/4}\\
e^{-i\pi/4}& -1
\end{array}\right],\label{fig_1}
\end{eqnarray}
is launched from site $0$ with initial coin state $\frac{\sqrt{2}}{2} |\downarrow\rangle +i\frac{\sqrt{2}}{2}|\uparrow\rangle$.
Then, by Corollary 1, we have 
\begin{eqnarray}
f(y)=\rho\delta(y)+\frac{(6+2\sqrt{2})y^2I_{(0,\sqrt{2}/2}(y)}{\pi (1-y^4)\sqrt{1-2y^2}}, 
\end{eqnarray}
where $\rho=\frac{(\sqrt{3}-\sqrt{2})(3-\sqrt{3})}{6}$.

The coefficient $\rho$ (see Eq. (\ref{weight_delta})) acts as the weight of the Dirac function $\delta(y)$ in the density function $f(y)$ and thereby governs the propensity for localization near the origin. In fact it is not difficult to see that $\rho$ may be computed as the sum of the limiting probabilities at all nodes, namely $\rho=\sum_{x=0}^{\infty}\lim_{t\rightarrow \infty}|\psi(x,t)|^2$. Accordingly, for the purposes of numerical simulation, it suffices to consider only probability values not very far removed from the initial position and time, say $x\leq 4$ and $t = 400$, as in Table 1:

\begin{table}[h]
\caption{\mbox{\footnotesize{Probability Distribution of $X_t$ for $x\leq 4$ and $t=400$}}} 
\centering  
\begin{tabular}{c c c ccc} 
\hline\hline                        
$x$ &0 &1 & 2 & 3 & 4  \\ [0.5ex] 
\hline                  
$|\psi(x,t)|^2$ & .0492& .0132& .0035 &.0010 &.0002  \\ 
\hline 
\end{tabular}
\label{table:nonlin} 
\end{table}

In terms of the tabulated values, we get $\rho\approxeq \sum_{x=0}^{4}|\psi(x,t)|^2\approxeq 0.0671\approxeq \frac{(\sqrt{3}-\sqrt{2})(3-\sqrt{3})}{6}.$ The limiting behavior of this specific QW is displayed in Figure 1. Note the localization spike near the origin. 
\begin{figure}[h]
\begin{center}
\includegraphics[height=2.5in,width=5in,angle=0]{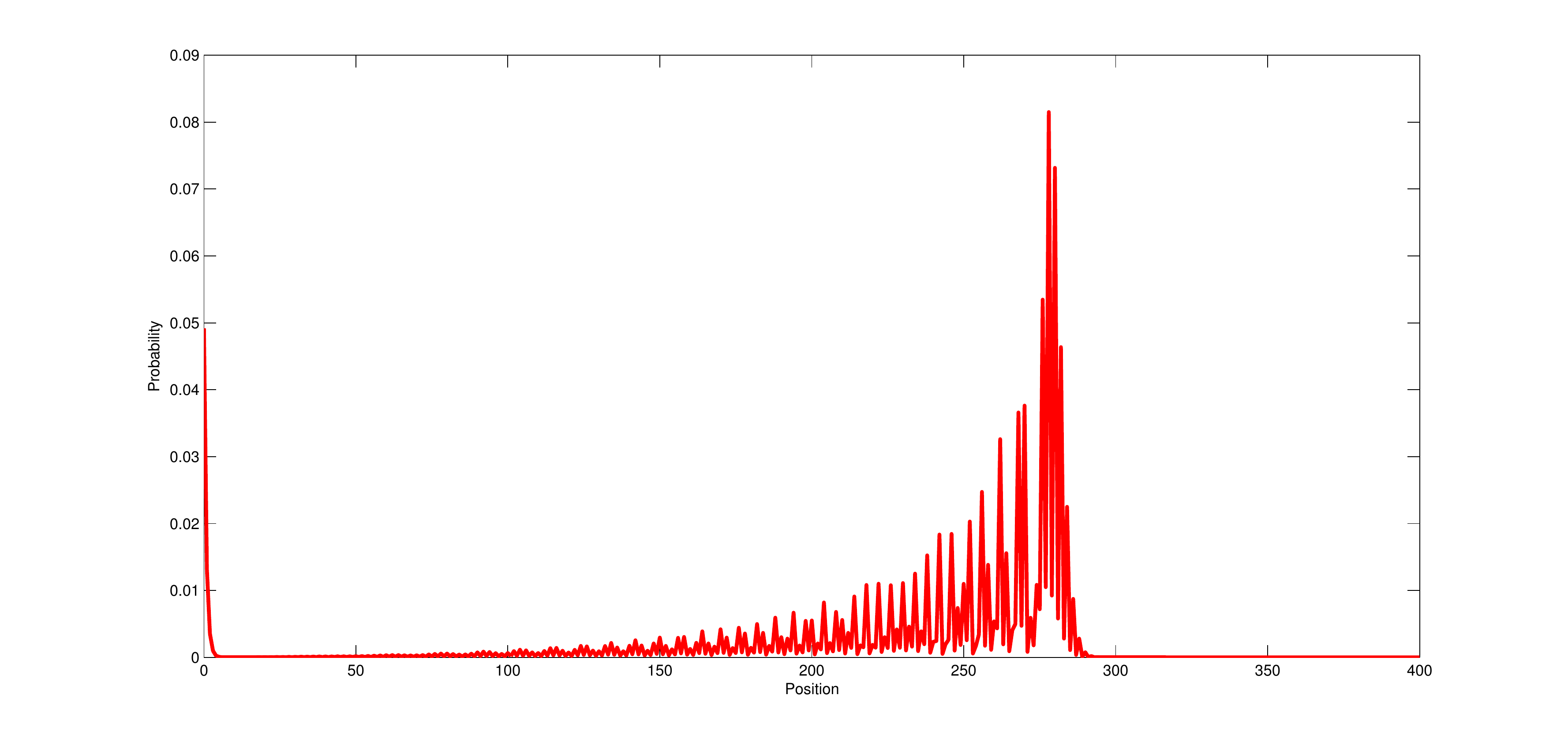}
\caption{\footnotesize{ Probability distribution at time $t=400$ of a homogeneous QW on the half-line with unitary coin operator $U$ as defined by Eq. (\ref{fig_1}) and initial coin state $\frac{\sqrt{2}}{2}|\downarrow\rangle+i\frac{\sqrt{2}}{2}|\uparrow\rangle$.}}
\end{center}
\end{figure}

However, as shown by our next example, localization is not always present for a homogeneous unitary QW on the half line. Let the Hadamard matrix $H$ (see Eq.(\ref{H})) serve as the constant coin operator for the homogeneous QW. By Corollary 1, we deduce:

{\bf Corollary 2.}\, Suppose the Hadamard QW is launched from site $0$ with the initial coin state $\alpha |\downarrow\rangle +\beta|\uparrow\rangle$, then

\begin{eqnarray}
X_t/t\Rightarrow Y
\end{eqnarray}
where $Y$ is a random variable with probability density 
$$f(y)=\frac{2 I_{(0,\sqrt{2}/2)}(y)}{\pi (1-y^2)\sqrt{1-2y^2}}.$$

For this QW, the probability distribution is depicted in Figure 2.

\begin{figure}[h]
\begin{center}
\includegraphics[height=2.5in,width=5in,angle=0]{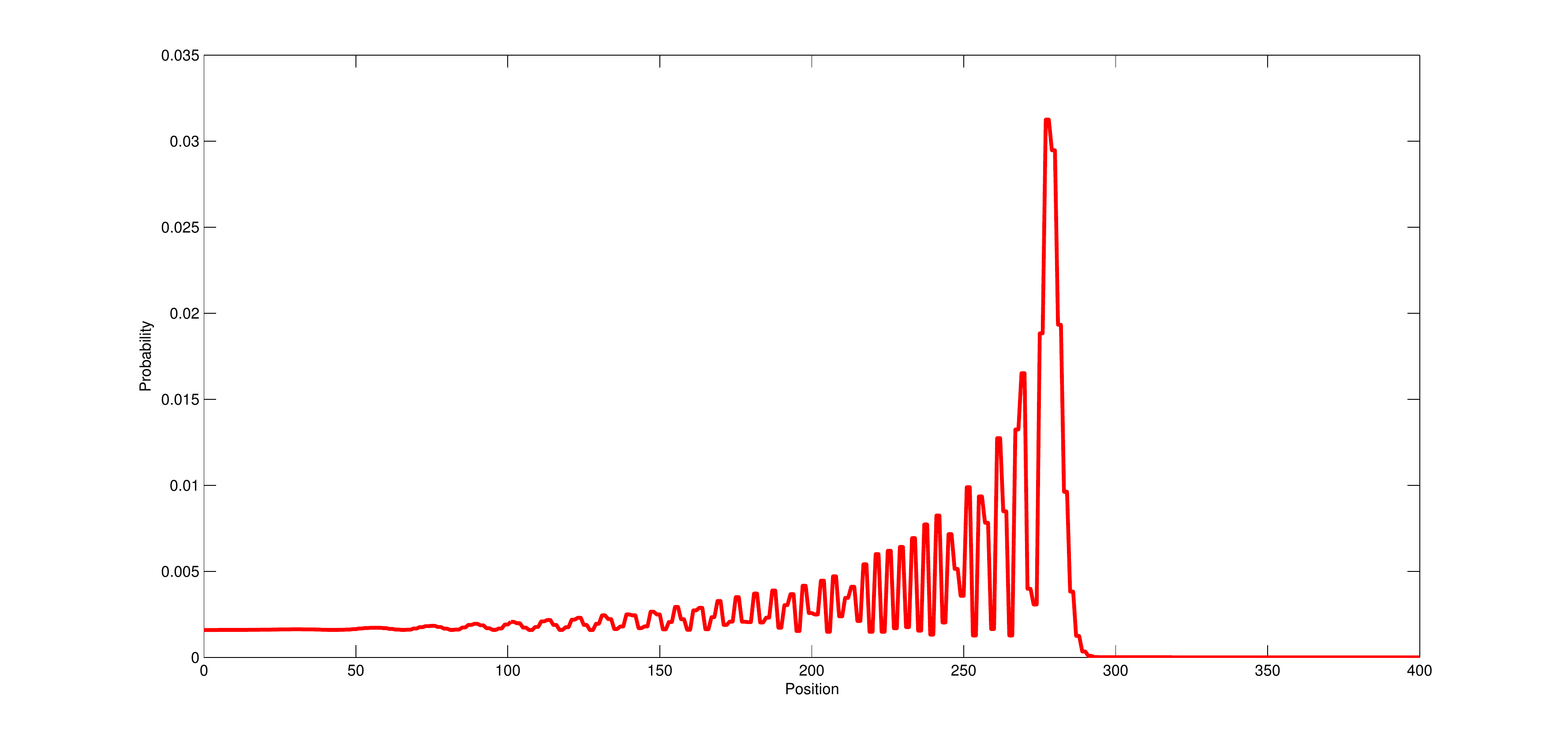}
\caption{\footnotesize{Probability distribution at time $t=400$ of a homogeneous QW on the half-line with Hadamard coin operator $H$ as defined by Eq.(\ref{H}) and initial coin state $\frac{\sqrt{2}}{2}|\downarrow\rangle+i\frac{\sqrt{2}}{2}|\uparrow\rangle$.}}
\end{center}
\end{figure}

In figure 1, the localization effect is evidenced by the egregious spike in the vicinity of the initial position. In figure 2, no such spike occurs near the initial position. The difference between the two graphs is completely determined by the value of the coefficient $\rho$.  In figure 1, since $\rho>0$, the height of the stationary spike near the initial position approximates the value of the limiting probability of finding the walker at that position, whereas, in figure 2, since $\rho=0$, no stationary spike is observed near the initial position.

Notably, according to Corollary 2, at least when the QW on the half-line is governed by the Hadamard coin operator, the weak limit is seen to be independent of the initial coin state.

Last, but not least, let us consider an example of an inhomogeneous quantum walk on the half line. 

{\bf Example 2.}\, Let the QW on the half line be launched with initial coin state $|\uparrow\rangle$ and coin operators $U$ and $U_{0}$   defined as follows:

\begin{equation}
 U_0= \frac{\sqrt{2}}{2}\left[\begin{array}{cc}
1& e^{i\pi/4}\\
e^{-i\pi/4}& -1
\end{array}\right],\quad
U=H= \frac{\sqrt{2}}{2}\left[\begin{array}{cc}
1& 1\\
1& -1
\end{array}\right]. \label{fig_2}
\end{equation}

By Theorem 1, the density for this QW is given by 

\begin{eqnarray}
f(y)
=\rho_{0}\delta(y) 
+\(\frac{k_{1}y^{2}+k_{2}y^4+k_{3}y^6}{m_{1}+m_{2}y^2+m_{3}y^4+m_{4}y^6}\)\frac{I_{(0,\sqrt{2}/2)}(y)}{\pi\sqrt{1-2y^2}},
\end{eqnarray}
where the numerical coefficients are
\begin{eqnarray}
\rho_{0}=0.677887, \,\,k_{1}=54-38\sqrt{2} , \,\,k_{2}=46\sqrt{2}-62 , \,\,k_{3}=16\sqrt{2}-16 \nonumber\\
m_{1}=17-12\sqrt{2} , \,\,m_{2}=64\sqrt{2}-90 , \,\,m_{3}=121-84\sqrt{2} , \,\,m_{4}= (4-4\sqrt{2})^{2}.\nonumber
\end{eqnarray}

The relatively large value of $\rho\approx 0.678$ indicates a strong tendency for localization near the origin. In Table 2 and Figure 3 we display numerical data for this QW at time $t=400$.

\begin{table}[h]
\caption{\footnotesize{Probability Distribution of $X_t$ for $x\leq 7$ and $t=400$}} 
\centering  
\begin{tabular}{c c c c c c c c c } 
\hline\hline                     
x &0 &1 & 2 & 3 & 4 &5 &6 &7  \\ [0.5ex] 
\hline                  
$|\psi(x,t)|^2$ & .4428& .1648& .0299&.0275 &.0046& .0054 &.0010 &.0012 \\ 
\hline 
\end{tabular}
\label{table:nonlin} 
\end{table}

Using the tabulated data, we verify that $\rho_{0}\approxeq \sum_{x=0}^{7}|\psi(x,t)|^2 \approxeq 0.6771.$

\begin{figure}[h]
\begin{center}
\includegraphics[height=2.5in,width=5in,angle=0]{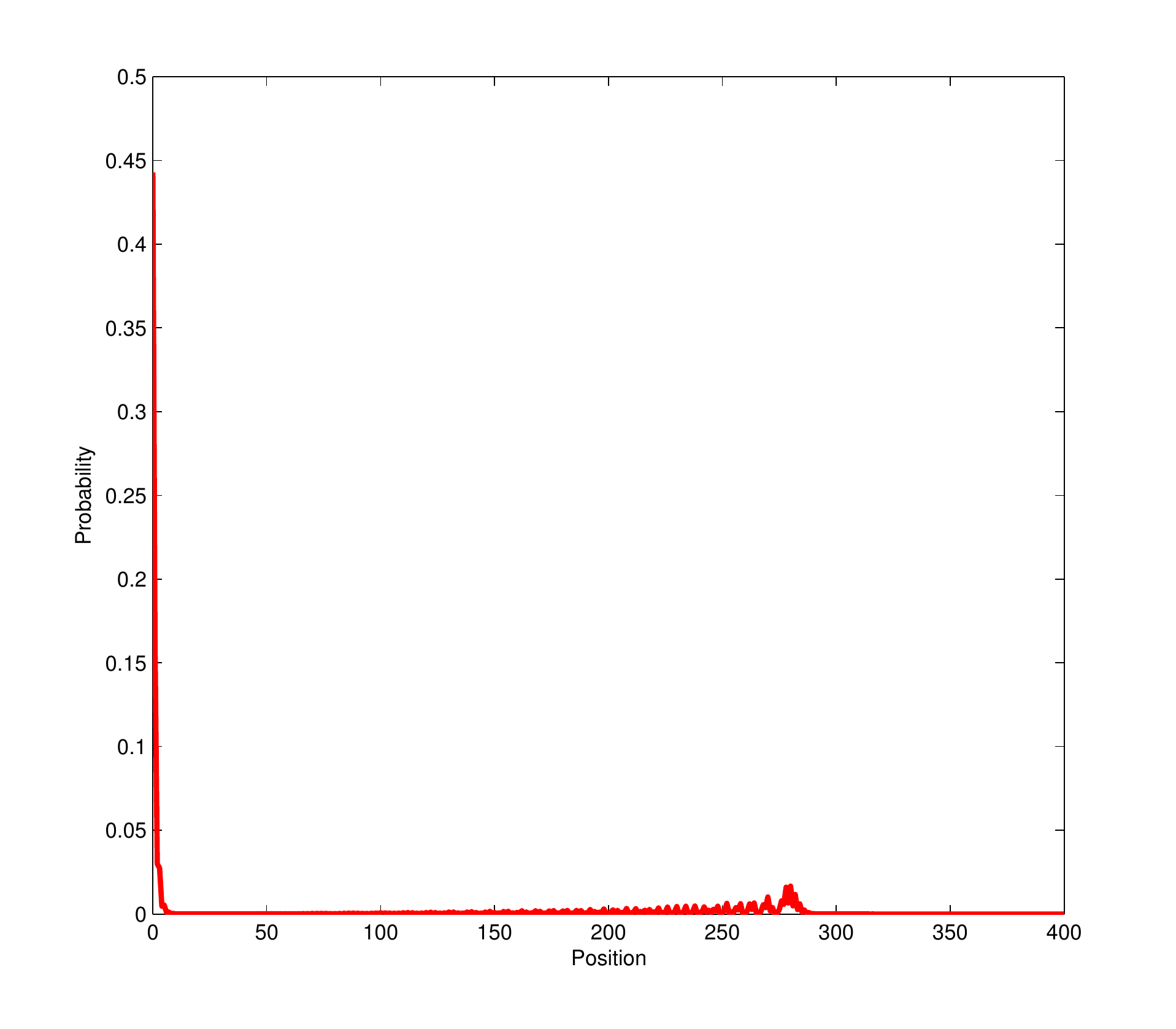}
\caption{\footnotesize{Probability distribution at time $t=400$ of a non-homogeneous QW on the half-line with coin operators $U_0$ and $U$ as defined by Eq. (\ref{fig_2}) and initial coin state $|\uparrow\rangle$.}}
\end{center}
\end{figure}

Arguably, in the literature, the publication most closely related to our present treatment of weak limits of QWs on the half-line is the investigation in \cite{CKS12} of a QW on a graph consisting of joined half lines. In that work, a similar weak limit is obtained for the graph. Other relevant works include \cite{K02, Konno05}, wherein weak limit results ("Konno's density") are derived for QWs on the full line. Numerous variations of the same theme can be found in the literature, including  \cite{KLS12, GJS04,IKS05, KFK05, MKK07, SK08, LP09, KM10, Liu12}, all of which exploit, to some degree, the
notion of Konno's density.

\section{Proof of Lemma 1}

To obtain the desired generating functions, we begin by introducing some convenient notations. Let $\Xi(x,t)$ denote the transition amplitude of the given QW, defined as the sum of the contributions from all paths starting at the site $0$ and ending at the site $x$ after $t$ steps. For the specific model of a QW with a breakdown at the site $0$, as in \cite{OKAA05}, let $\Xi^b(x,t)$ denote $\Xi^b(0\rightarrow x,t)$ as in  \cite{OKAA05}. To denote the generating functions $B^r(0\rightarrow 0,z)$, $B^q(0\rightarrow x,z)$ and $B^r(0\rightarrow x,z)$, as defined in \cite{OKAA05}, the abbreviated symbols $B^r(0,z)$, $B^q(x,z)$ and $B^r(x,z)$ respectively are introduced. 


In addition to the special matrices $S_{0},\,\, Q_{0},\,\, P\,\, \mbox{and}\,\, Q$ defined in Eq. (\ref{matrices}), let
\begin{equation}
P_0= \left[\begin{array}{cc}
a_0& b_0\\
0& 0
\end{array}\right],
R_0= \left[\begin{array}{cc}
c_0& d_0\\
0& 0
\end{array}\right],
S= \left[\begin{array}{cc}
0& 0\\
a& b
\end{array}\right],
R= \left[\begin{array}{cc}
c& d\\
0& 0
\end{array}\right]
\end{equation}

By the $PQRS$ method \cite{K02}, the transition amplitudes can be expressed as unique linear combinations of the matrices $P_0, Q_0, R_0$ and $S_0$ (or $P, Q, R$ and $S$):
\begin{eqnarray}
\Xi(x,t)=u^{p_0}(x,t)P_0+u^{q_0}(x,t)Q_0+u^{r_0}(x,t)R_0+u^{s_0}(x,t)S_0.\label{linear-expression}
\end{eqnarray}
 
The expressions in Eq. (\ref{linear-expression}) can be represented as coefficients of the following four generating functions in $z$:

\begin{eqnarray}
U^{p_0}(x,z)=\sum_{t=0}^{\infty}u^{p_0}(x,t)z^t,\, U^{q_0}(x,z)=\sum_{t=0}^{\infty}u^{q_0}(x,t)z^t\\
U^{r_0}(x,z)=\sum_{t=0}^{\infty}u^{r_0}(x,t)z^t,\, U^{s_0}(x,z)=\sum_{t=0}^{\infty}u^{s_0}(x,t)z^t.
\end{eqnarray}

Now by Eq. (\ref{recursive-amplitudes}), the transition amplitudes at $x=1$ are given by:
\begin{eqnarray}
\!\!\!\Xi(1,1)
\!\!&=&\!\!Q_0 \nonumber\\
\!\!\!\Xi(1,2)
\!\!&=&\!\!Q_0S_0=d_0S_0 \nonumber\\
\!\!\!\Xi(1,t)
\!\!&=&\!\!\sum_{l}\Xi^b(0,t-1-l)Q_0\Xi(0,l) \nonumber\\
\!\!&+&\!\!\Xi(0,t-1)Q_0+Q_0\Xi(0,t-1)\, \mathrm{for}\, t\ge 3 \label{transition-1}
\end{eqnarray}

in terms of which, we obtain the following formulas for the generating functions at $x=1$.

\begin{eqnarray}
\!\!\!U^{p_0}(1,z)
\!\!&=&\!\![c_0U^{p_0}(0,z)+d_0U^{s_0}(0,z)]dzB^r(0,z),\nonumber\\
\!\!\!U^{q_0}(1,z)
\!\!&=&\!\!z+d_0zU^{q_0}(0,z)+c_0zU^{r_0}(0,z),\nonumber\\
\!\!\!U^{r_0}(1,z)
\!\!&=&\!\!\![1+d_0U^{q_0}(0,z)+c_0U^{r_0}(0,z)]dzB^r(0,z),\nonumber\\
\!\!\!U^{s_0}(1,z)
\!\!&=&\!\!c_0zU^{p_0}(0,z)+d_0zU^{s_0}(0,z). \label{function-1}
\end{eqnarray}

Again by Eq. (\ref{recursive-amplitudes}), we have: 

\begin{eqnarray}
P\Xi(1,t)+S_0\Xi(0,t)=\Xi(0,t+1) \label{relation-01}
\end{eqnarray}

Also, by Eqs. (\ref{relation-01}) and (\ref{function-1}) we obtain the following formulas for the generating functions at $x=0$:

\begin{eqnarray}
\!\!\!U^{p_0}(0,z)
\!\!&=&\!\!\frac{bd_0z^3+add_0z^3B^r(0,z)}{1-b_0z-bc_0z^2-b\Delta_0z^3-ad\Delta_0z^3B^r(0,z)-ac_0dz^2B^r(0,z)}, \nonumber\\
\!\!\!U^{q_0}(0,z)
\!\!&=&\!\!\frac{a_0bz^3+aa_0dz^3B^r(0,z)}{1-b_0z-bc_0z^2-b\Delta_0z^3-ad\Delta_0z^3B^r(0,z)-ac_0dz^2B^r(0,z)}, \nonumber\\
\!\!\!U^{r_0}(0,z)
\!\!&=&\!\!\frac{(1-b_0z)[bz^2+adz^2B^r(0,z)]}{1-b_0z-bc_0z^2-b\Delta_0z^3-ad\Delta_0z^3B^r(0,z)-ac_0dz^2B^r(0,z)}, \nonumber\\
\!\!\!U^{s_0}(0,z)
\!\!&=&\!\!\frac{z-bc_0z^3-ac_0dz^3B^r(0,z)}{1-b_0z-bc_0z^2-b\Delta_0z^3-ad\Delta_0z^3B^r(0,z)-ac_0dz^2B^r(0,z)} \label{function-0}
\end{eqnarray}

When $x\ge 2$, Eq. (\ref{recursive-amplitudes}) yields:

\begin{eqnarray}
\!\!\!\Xi(x,x)
\!\!&=&\!\!\Xi^b(x-1,x-1)Q_0 \nonumber\\
\!\!\!\Xi(x,t)
\!\!&=&\!\!\sum_{0<l\le t-x}\Xi^b(x-1,t-1-l)Q_0\Xi(0,l) \nonumber\\
\!\!&+&\!\!\Xi^b(x-1,t-1)Q_0 \, \mathrm{for}\, t\ge x+1, \label{transition-x}
\end{eqnarray}

and subsequently:

\begin{eqnarray}
\!\!\!U^{p_{0}}(x,z)
\!\!&=&\!\![c_0U^{p_0}(0,z)+d_0U^{s_0}(0,z)]dz B^{r}(x-1,z), \nonumber\\
\!\!\!U^{q_0}(x,z)
\!\!&=&\!\![1+d_0 U^{q_0}(0,z)+c_0 U^{r_0}(0,z)]dz B^q(x-1,z), \nonumber\\
\!\!\!U^{r_0}(x,z)
\!\!&=&\!\![1+d_0U^{q_0}(0,z)+c_0U^{r_0}(0,z)]dz B^r(x-1,z), \nonumber\\
\!\!\!U^{s_0}(x,z)
\!\!&=&\!\![c_0U^{p_0}(0,z)+d_0U^{s_0}(0,z)]dz B^q(x-1,z). \label{function-x}
\end{eqnarray}

Finally, we have:

\begin{eqnarray}
\!\!\!\Psi_{\downarrow}(0,z)
\!\!&=&\!\!(1,0)\psi(0,0)+\sum_{t=1}^{\infty}(1,0)\Xi(x,t)\psi(0,0)z^t \nonumber\\
\!\!&=&\!\!\alpha+(\alpha c_0+\beta d_0)U^{r_0}(0,z)+(\alpha a_0+\beta b_0)U^{p_0}(0,z) \nonumber\\
\!\!\!\Psi_{\downarrow}(x,z)
\!\!&=&\!\!\sum_{t=1}^{\infty}(1,0)\Xi(x,t)\psi(0,0)z^t \nonumber\\
\!\!&=&\!\!(\alpha c_0+\beta d_0)U^{r_0}(x,z)+(\alpha a_0+\beta b_0)U^{p_0}(x,z) \nonumber\\
\!\!\!\Psi_{\uparrow}(0,z)
\!\!&=&\!\!(0,1)\psi(0,0)+\sum_{t=1}^{\infty}(0,1)\Xi(x,t)\psi(0,0)z^t \nonumber\\
\!\!&=&\!\!\beta+(\alpha c_0+\beta d_0)U^{q_0}(0,z)+(\alpha a_0+\beta b_0)U^{s_0}(0,z) \nonumber\\
\!\!\!\Psi_{\uparrow}(x,z)
\!\!&=&\!\!\sum_{t=1}^{\infty}(0,1)\Xi(x,t)\psi(0,0)z^t \nonumber\\
\!\!&=&\!\!(\alpha c_0+\beta d_0)U^{q_0}(x,z)+(\alpha a_0+\beta b_0)U^{s_0}(x,z) \label{definitions}
\end{eqnarray}

Together, Eqs.({\ref{function-1}), (\ref{function-0}), (\ref{function-x}) and (\ref{definitions}) suffice to complete the proof.

\section {Proof of Theorem 1}

Consider the Fourier transform of the generating function $\Psi(x,z)$, which is given by $\widehat{\Psi}(k,z)=\sum_{x=0}^{\infty}\Psi(x,z)e^{ixk}$. Collecting like terms in $z^t$, we get $\widehat{\Psi}(k,z)=\sum_{t=0}^{\infty}\Psi(k,t)z^t$ where $\Psi(k,t)=\sum_{t=0}^{\infty}\psi(x,t)e^{ixk}$. By Lemma 1, one has a closed formula for $\widehat{\Psi}(k,z)$, by which it can be shown that $\widehat{\Psi}(k,z)$ has two types of unit poles: the first type of poles are the zeros (assumed to be $\{\omega_j: j\in J \}$) of $1-b_0z-bc_0z^2-b\Delta_0z^3-ad\Delta_0z^3B^r(0,z)-ac_0dz^2B^r(0,z)$, which contribute to the part of a mass point represented by the distribution $\rho\delta(y)$ in the formula for $f(y)$; while the second type of pole, as the zero of $1-\frac{d}{a}\lambda_{+}e^{ik}$, determines the absolutely continuous part of the density $f(y)$. Let us introduce $a\Delta^{-\frac{1}{2}}=|a|e^{i\eta}$ with $0\le \eta <2\pi$, then this single pole is $e^{i\theta_1(k)}$ when $k\in (\eta-\pi,\eta)$, or $e^{i\theta_2(k)}$ when $k\in (\eta-2\pi,\eta-\pi)$. Here $\theta_1(k)=\arccos(|a|\cos(\eta-k))$ and $\theta_2(k)=2\pi-\arccos(|a|\cos(\eta-k))$.

 
By Cauchy's integral theorem, 

\begin{eqnarray}
   \!\!\!\Psi(k,t)
\!\!&=&\!\!\frac{1}{2\pi i}\oint_{\{z:|z|=r\, \mathrm{for}\,0<r<1\}} \frac{\widehat{\Psi}(k,z)}{z^{t+1}}dz\\
 \!\!&=&\!\! \left\{
     \begin{array}{lr}
    -\mathrm{Res}(\frac{\widehat{\Psi}(k,z)}{z^{t+1}}, e^{i\theta_1})-\sum_{j\in J}\mathrm{Res}(\frac{\widehat{\Psi}(k,z)}{z^{t+1}}, \omega_j), & k\in (\eta-\pi,\eta)  ,\\
      -\mathrm{Res}(\frac{\widehat{\Psi}(k,z)}{z^{t+1}}, e^{i\theta_2})-\sum_{j\in J}\mathrm{Res}(\frac{\widehat{\Psi}(k,z)}{z^{t+1}}, \omega_j) , & k\in (\eta-2\pi,\eta-\pi).
     \end{array}
   \right.
\end{eqnarray}



By the structure of the function $\Psi(k,t)$, one can calculate the characteristic function of the scaled quantity $X_t/t$ so that 
\begin{eqnarray}
\!\!\mathrm{E}(e^{i\xi X_t/t})
\!\!&=&\!\!\int_{\eta-2\pi}^{\eta}\langle \Psi(k,t),\Psi(k+\xi/t,t)\rangle \frac{dk}{2\pi} \nonumber\\
\!\!&=&\!\!\int_{\eta-\pi}^{\eta}e^{i(t+1)[\theta_1(k)-\theta_1(k+\xi/t)]}\overline{\mathrm{Res}(\widehat{\Psi}(k,z),e^{i\theta_1(k)})}\mathrm{Res}(\widehat{\Psi}(k+\xi/t,z),e^{i\theta_1(k+\xi/t)})\frac{dk}{2\pi} \nonumber\\
\!\!&+&\!\!\int_{\eta-2\pi}^{\eta-\pi}e^{i(t+1)[\theta_2(k)-\theta_2(k+\xi/t)]}\overline{\mathrm{Res}(\widehat{\Psi}(k,z),e^{i\theta_2(k)})}\mathrm{Res}(\widehat{\Psi}(k+\xi/t,z),e^{i\theta_2(k+\xi/t)})\frac{dk}{2\pi}\nonumber\\
\!\!&+&\!\!\int_{\eta-2\pi}^{\eta}\overline{\sum_{j\in J}\mathrm{Res}(\frac{\widehat{\Psi}(k,z)}{z^{t+1}},\omega_j)}\sum_{j\in J}\mathrm{Res}(\frac{\widehat{\Psi}(k+\xi/t,z)}{z^{t+1}},\omega_j)\frac{dk}{2\pi} \nonumber\\
\!\!&+&\!\!\int_{\eta-\pi}^{\eta}e^{i(t+1)\theta_1(k)}\overline{\mathrm{Res}(\widehat{\Psi}(k,z),e^{i\theta_1(k)})}\sum_{j\in J}\mathrm{Res}(\frac{\widehat{\Psi}(k+\xi/t,z)}{z^{t+1}},\omega_j)\frac{dk}{2\pi} \nonumber\\
\!\!&+&\!\!\int_{\eta-\pi}^{\eta}e^{-i(t+1)\theta_1(k+\xi/t)}\mathrm{Res}(\widehat{\Psi}(k+\xi/t,z),e^{i\theta_1(k+\xi/t)})\overline{\sum_{j\in J}\mathrm{Res}(\frac{\widehat{\Psi}(k,z)}{z^{t+1}},\omega_j)}\frac{dk}{2\pi} \nonumber\\
\!\!&+&\!\!\int_{\eta-2\pi}^{\eta-\pi}e^{i(t+1)\theta_2(k)}\overline{\mathrm{Res}(\widehat{\Psi}(k,z),e^{i\theta_2(k)})}\sum_{j\in J}\mathrm{Res}(\frac{\widehat{\Psi}(k+\xi/t,z)}{z^{t+1}},\omega_j)\frac{dk}{2\pi} \nonumber\\
\!\!&+&\!\!\int_{\eta-2\pi}^{\eta-\pi}e^{-i(t+1)\theta_2(k+\xi/t)}\mathrm{Res}(\widehat{\Psi}(k+\xi/t,z),e^{i\theta_2(k+\xi/t)})\overline{\sum_{j\in J}\mathrm{Res}(\frac{\widehat{\Psi}(k,z)}{z^{t+1}},\omega_j)}\frac{dk}{2\pi} \label{cha-X}
\end{eqnarray}


As $t\rightarrow\infty$, each of the last four terms in Eq. (\ref{cha-X}) converges to 0 by Riemann-Lebesgue lemma, and the third term converges to $\sum_{x=0}^{\infty}\lim_{t\rightarrow \infty}|\psi(x,t)|^2$, which is $\rho$ in the formula for the density function $f(y)$. Therefore we get

\begin{eqnarray}
\!\!\!\lim_{t\rightarrow \infty}E(e^{i\xi X_t/t}) 
\!\!&=&\!\!\int_{\eta-\pi}^{\eta}e^{-i\xi \theta_1^{\prime}(k)}\|\mathrm{Res}(\widehat{\Psi}(k,z),e^{i\theta_1(k)})\|^2\frac{dk}{2\pi} \nonumber\\
\!\!&+&\!\!\int_{\eta-2\pi}^{\eta-\pi}e^{-i\xi \theta_2^{\prime}(k)}\|\mathrm{Res}(\widehat{\Psi}(k,z),e^{i\theta_2(k)})\|^2\frac{dk}{2\pi}\nonumber\\
\!\!&+&\!\!\rho \label{y-density}
\end{eqnarray}
where $\theta_1^{\prime}(k)=-\frac{|a|\sin(\eta-k)}{\sqrt{1-|a|^2\cos^2(\eta-k)}}$ and $\theta_2^{\prime}(k)=\frac{|a|\sin(\eta-k)}{\sqrt{1-|a|^2\cos^2(\eta-k)}}$.
 
Let $y=|\frac{|a|\sin(\eta-k)}{\sqrt{1-|a|^2\cos^2(\eta-k)}}|$. After an extensive calculation, Eq. (\ref{y-density}) reduces to the identity:
\begin{eqnarray}
\lim_{t\rightarrow \infty}E(e^{i\xi X_t/t})=\rho+\int_0^{|a|}e^{i\xi y}h(y)\frac{|c|^2y^2}{\pi (1-y^2)\sqrt{|a|-y^2}}dy \label{fdensity}
\end{eqnarray}
with $h(y)$ is given by Eqs. (\ref{h(y)1}) and (\ref{h(y)2}). Setting $\xi=0$ in Eq. (\ref{fdensity}) yields Eq. (\ref{weight_delta}), whereby the proof is complete.

{\bf  Acknowledgments}

 Undergraduate research assistant and mathematics major Forrest Ingram-Johnson is to be commended for performing the computer simulations for the graphs in Fig. 1-3. Liu was supported, in part, by NSF Grant CCF-1005564.



\begin{thebibliography}{0}

\bibitem{ADZ93}

Y. Aharonov, L. Davidovich, and N. Zagury, Phys. Rev. A {\bf 48}, 1687 (1993).

\bibitem{ABNVW01}

A. Ambainis, E. Bach, A. Nayak, A. Vishwanath and J. Watrous, {\it One-dimensional quantum walks},
in Proceedings of the thirty-third annual ACM symposium on Theory of computing, STOC '01 (ACM, New York, NY, USA,
2001) pp. 37-49.

\bibitem{AAKV01}

D. Aharanov, A. Ambainis, J. Kempe and U. Vazirani, {\it Quantum walks on graphs}, in Proceedings of the 33rd Annual ACM Symposium on Theory of Computing, (ACM, New York, 2001), pp.50-59.

\bibitem{FG98}


E. Farhi, S. Gutmann, {\it Quantum computation and decision trees}, Phys. Rev. A 58, 915–928 (1998).


\bibitem{K03}

J. Kempe (2003), {\it Quantum random walks - an introductory overview}, Contemp. Phys. {\bf 44}, 307.

\bibitem{K06}

 V. Kendon, {\it Decoherence in quantum walks - a review}, Struct. in Comp. Sci 17(6) pp 1169-1220 (2006).

\bibitem{K08}

N. Konno, in {\em Quantum Potential Theory}, Lecture Notes in Mathematics, edited by U. Franz and M. Schurmann (Springer-Verlag, Heidelberg, 2008), pp.309-452.

\bibitem{VA08}

S.E. Venegas-Andraca, {\em Quantum Walks for Computer Scientists}, Morgan and Claypool Publishers (Synthesis Lectures on Quantum Computing), 2008.

\bibitem{AMC2009}

A. M. Childs, {\it Universal computation by quantum walk}, Phys. Rev. Lett. 102, 180501 (2009).

\bibitem{AVWW11}

A. Ahlbrecht, H. Vogts, A. H. Werner, and R. F. Werner, {\it Asymptotic evolution of quantum walks with random coin}, J. Math. Phys. 52, 042201 (2011).

\bibitem{VA12}

S.E. Venegas-Andraca, {\em Quantum walks: a comprehensive review}, Quantum Information Processing vol. 11(5), pp. 1015-1106 (2012).

\bibitem{CGMV2012}

M. J. Cantero, F. A. Gr$\ddot{\mathrm{u}}$nbaum, L. Moral and L. Vel$\acute{\mathrm{a}}$zquez (2012), {\it The CGMV method for quantum walks},Quantum Inf Process (2012) 11:1149–1192.


\bibitem{K02}

N. Konno, {\it Quantum random walks in one dimension}, Quantum Information Processing, 1:, pp. 345-354 (2002).

\bibitem{Konno05}
N. Konno, {\it A new type of limit theorems for the one-dimensional quantum random walk},
Journal of the Mathematical Society of Japan, 57: 1179-1195 (2005).

\bibitem{KLS12}
N. Konno, T. Luczak, and E. Segawa, {\it Limit measures of inhomogeneous discrete-time quantum walks in one dimension},Quantum Inf Process (2013) 12:33–53.


\bibitem{CGMV2010}

M. J. Cantero, F. A. Gr$\ddot{\mathrm{u}}$nbaum, L. Moral and L. Vel$\acute{\mathrm{a}}$zquez (2010), {\it Matrix-valued Szeg$\ddot{o}$ polynomials and quantum random walks}, Communications on Pure and Applied Mathematics, 63, 464-507.

\bibitem{CGMV1-2012}

M. J. Cantero, F. A. Gr$\ddot{\mathrm{u}}$nbaum, L. Moral and L. Vel$\acute{\mathrm{a}}$zquez (2012), {\it One-dimensional quantum walks with one
defect}. Rev. Math. Phys. 24, 1250002.


\bibitem{CHKS2009}

K. Chisaki, M. Hamada, N. Konno, E. Segawa (2009), {\it Limit theorems for discrete-time quantum walks on trees},
Interdiscip. Inform. Sci. 15, 423-429.




\bibitem{OKAA05}

Oka, T., Konno, N., Arita, R., and Aoki, H., {\it Breakdown of an electric-field driven system: a
mapping to a quantum walk}, Physical Review Letter, 94: 100602 (2005).



\bibitem{CKS12}

K. Chisaki, N. Konno, and E. Segawa, {\it Limit Theorems for the Discrete-Time Quantum Walk on a Graph with Joined Half Lines}, Quantum Information and Computation 12, (2012) 0314-0333.



\bibitem{GJS04}

G. Grimmett, S. Janson and P. F. Scudo (2004), {\it Weak limits for quantum random walks}, Phys. Rev. E {\bf 69}, 026119.



\bibitem{IKS05}
N.Inui, N. Konno and E. Segawa,{\it One-dimensional three-state quantum walk}, Phys. Rev. E 72 (2005) 056112.

\bibitem{KFK05}
M. Katori, S. Fujino and N. Konno, {\it  Quantum walks and orbital states of a Weyl particle}. Phys. Rev. A
72 (2005) 012316.

\bibitem{MKK07}
T. Miyazaki, M.  Katori  and N. Konno, {\it Wigner formula of rotation matrices and quantum walks}. Phys.
Rev. A 76 (2007) 012332.

\bibitem{SK08}
E. Segawa, N. Konno, {\it Limit theorems for quantum walks driven by many coins}. Int. J. Quantum Inf.
6 1231 (2008).

\bibitem{LP09}
C. Liu, N. Petulante, {\it One-dimensional quantum random walk with two entangled coins}. Phy. Rev. A
79 032312 (2009).

\bibitem{KM10}
N. Konno, T. Machida, {\it Limit theorems for quantum walks with memory}. Quant. Inf. Comput. 10
1004 (2010).

\bibitem{Liu12}

C. Liu, {\it Asymptotic distributions of quantum walks on the line with two entangled coins}, Quantum Inf Process (2012) 11:1193-1205.

\end{thebibliography}
\end{document}